\documentclass[
superscriptaddress,
twocolumn,
showpacs,preprintnumbers,
amsmath,amssymb,
aps,
prb,
]{revtex4-2}

\usepackage[colorlinks,
linkcolor=blue, urlcolor=blue, anchorcolor=blue, citecolor=blue]{hyperref}

\usepackage{graphicx}
\usepackage{dcolumn}
\usepackage{bm}
\usepackage{float}
\usepackage{xcolor}

\newcommand{\angstrom}{\textup{\AA}}

\begin{document}


\title{Triaxial magnetic anisotropy in the two-dimensional ferromagnetic semiconductor CrSBr}


\author{Ke Yang}
\affiliation{College of Science, University of Shanghai for Science and Technology, Shanghai 200093, China}
 \affiliation{Laboratory for Computational Physical Sciences (MOE),
 State Key Laboratory of Surface Physics, and Department of Physics,
 Fudan University, Shanghai 200433, China}

\author{Guangyu Wang}
 \affiliation{Laboratory for Computational Physical Sciences (MOE),
 State Key Laboratory of Surface Physics, and Department of Physics,
 Fudan University, Shanghai 200433, China}

\author{Lu Liu}
 \affiliation{Laboratory for Computational Physical Sciences (MOE),
 State Key Laboratory of Surface Physics, and Department of Physics,
 Fudan University, Shanghai 200433, China}

\author{Di Lu}
 \affiliation{Laboratory for Computational Physical Sciences (MOE),
 State Key Laboratory of Surface Physics, and Department of Physics,
 Fudan University, Shanghai 200433, China}

\author{Hua Wu}
\email{Corresponding author. wuh@fudan.edu.cn}
\affiliation{Laboratory for Computational Physical Sciences (MOE),
 State Key Laboratory of Surface Physics, and Department of Physics,
 Fudan University, Shanghai 200433, China}
\affiliation{Shanghai Qi Zhi Institute, Shanghai 200232, China}
\affiliation{Collaborative Innovation Center of Advanced Microstructures,
 Nanjing 210093, China}

%
%

\date{\today}

\begin{abstract}

Two-dimensional (2D) ferromagnets have recently drawn extensive attention, and here we study the electronic structure and magnetic properties of the bulk and monolayer of CrSBr, using first-principles calculations and Monte Carlo simulations. Our results show that bulk CrSBr is a magnetic semiconductor and has the easy magnetization $b$-axis, hard $c$-axis, and intermediate $a$-axis. Thus, the experimental triaxial magnetic anisotropy (MA) is well reproduced here, and it is identified to be the joint effects of spin-orbit coupling (SOC) and magnetic dipole-dipole interaction. We find that bulk CrSBr has a strong ferromagnetic (FM) intralayer coupling but a marginal interlayer one. We also study CrSBr monolayer in detail and find that the intralayer FM exchange persists and the shape anisotropy has a more pronounced contribution to the MA. Using the parameters of the FM exchange and the triaxial MA, our Monte Carlo simulations show that CrSBr monolayer has Curie temperature \textit{T}$_{\rm C}$ = 175 K. Moreover, we find that a uniaxial tensile (compressive) strain along the $a$ ($b$) axis would further increase the \textit{T}$_{\rm C}$.

\end{abstract}

\maketitle


\section*{I. Introduction}

There has been a tremendous interest in the exfoliatable van der Waals (vdW) magnetic materials since the recent discoveries of 2D FM in the atomically thin CrI$_{3}$~\cite{Huang2017} and Cr$_{2}$Ge$_{2}$Te$_{6}$ ~\cite{Gong2017}.
The FM ordering is remarkably retained in CrI$_{3}$ monolayer with the Curie temperature ($T_{\rm C}$) $\sim$ 45 K and in Cr$_{2}$Ge$_{2}$Te$_{6}$ bilayer with $T_{\rm C}$ $\sim$ 30 K under a tiny magnetic field. These discoveries stimulate extensive research on the emergent 2D magnetism, which has promising spintronic applications~\cite{Li2019,Song2019,Gong2019,NSR,Huang2018,Lado2017,Kim2019,Xu2018,Webster2018,Wilson2021}.
As the 2D isotropic Heisenberg spin systems have no long-range magnetic order at finite temperature according to the Mermin-Wagner theorem~\cite{MW}, magnetic anisotropy (MA) is indispensable for stabilizing the 2D magnetic order. Both CrI$_{3}$ monolayer~\cite{Huang2017} and Cr$_{2}$Ge$_{2}$Te$_{6}$ bilayer~\cite{Gong2017} have an easy out-of-plane magnetization, while the 2D magnet CrCl$_{3}$ shows an easy in-plane magnetization~\cite{Wang2019,Klein2019}. Understanding and exploration of the variable MA are vitally important for the 2D magnets~\cite{Lado2017, Kim2019, Xu2018, Yang2020, Liu2020, Ni2021, Sears2020}, and they would facilitate development of spintronic materials and devices.

\begin{figure}[t]
\includegraphics[width=7cm]{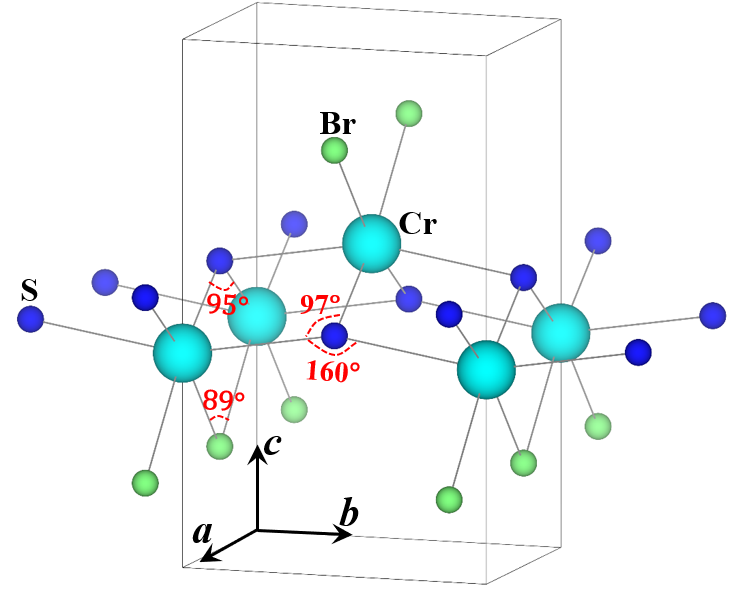}
 \caption{The bulk vdW crystal structure of CrSBr with the distorted CrS$_4$Br$_2$ octahedra.}
\label{Structure}
\end{figure}

CrSBr is a 2D magnetic material and has a vdW layered structure along the $c$-axis~\cite{Telford2020}, see FIG. \ref{Structure}. Its main structural blocks are the distorted CrS$_4$Br$_2$ octahedra, and they are connected by the SBr-edge sharing along the $a$-axis, by the S-corner sharing along the $b$-axis, and by the SS-edge sharing along the diagonal of the $ab$ plane, thus forming the 2D magnetic lattice. CrSBr bulk material is a layered antiferromagnetic (AF) semiconductor (intralayer FM and interlayer AF) with the N\'eel temperature $T_{\rm N}$ = 132 K and the optical band gap of about 1.25 eV~\cite{Telford2020}. The bulk CrSBr has a triaxial MA with the easy magnetization $b$-axis, the intermediate $a$-axis and the hard $c$-axis~\cite{Telford2020}. Several theoretical studies predict CrSBr monolayer to be a 2D FM with $T_{\rm C}$ of 160 $\sim$ 180 K~\cite{Jiang2018,Wang2020,Chen2020,Guo2018}. Experimentally, the $T_{\rm C}$ was measured to be 146 K for CrSBr monolayer very recently by the second harmonic generation technique~\cite{Monolayer_CrSBr}. In contrast to the well studied intralayer FM coupling and the 2D FM order~\cite{Jiang2018,Wang2020,Chen2020,Guo2018,Monolayer_CrSBr}, the triaxial MA~\cite{Telford2020} received much less attention previously, which is however indispensable for the 2D magnetic order. Moreover, compared with an SOC induced MA, a shape anisotropy due to the magnetic dipole-dipole interaction is often relatively weak and of less concern. But for a 2D system with a moderate or weak MA, the shape anisotropy may have an important contribution~\cite{Xue2019}, which is indeed the case for the CrSBr monolayer as demonstrated below.

In this work, we study the electronic structure and magnetic properties of the bulk and monolayer of CrSBr, using first-principles calculations and Monte Carlo simulations. Our results show that CrSBr bulk is indeed a magnetic semiconductor and has a charge transfer band gap, and that it has a strong FM intralayer coupling but a marginal interlayer one. It is important to note that our calculations well reproduce the experimental triaxial MA, which we identify to be the joint effects of the SOC and the magnetic dipole-dipole interaction. Moreover, we find that the CrSBr monolayer has a more pronounced shape anisotropy due to the magnetic dipole-dipole interaction, and that its robust intralayer FM coupling and the triaxial MA yield the $T_{\rm C}$ = 175 K based on our Monte Carlo simulations. Furthermore, we study the strain effects on the intralayer FM exchange and the MA of the CrSBr monolayer, and we find that a uniaxial tensile strain along the $a$-axis or a compressive strain along the $b$-axis would further increase the $T_{\rm C}$.

\section*{II. Computational Details}

We perform density functional theory calculations using the Vienna \textit{ab} initio simulation package (VASP)~\cite{VASP}.
The exchange-correlation effect is described by the generalized gradient approximation (GGA), using the functional proposed by Perdew, Burke, and Ernzerhof (PBE)~\cite{PBE}.
The vdW corrections are used in CrSbr bulk calculations within Grimme’s approach (DFT-D2)~\cite{vdW}. For CrSBr monolayer, the vacuum space between adjacent slabs is set to be 15 $\angstrom$.
We use the experimental bulk lattice constants $a$ = 3.50 \AA, $b$ = 4.76 \AA, and $c$ = 7.96 \AA~\cite{Telford2020} and optimize the atomic positions. The total energies and atomic forces converge to 0.1 $\mu$eV/fu and 1 meV/$\angstrom$, respectively. For CrSBr monolayer, the optimized lattice constants $a$ = 3.54 \AA, and $b$ = 4.73 \AA~ are almost the same as the experimental bulk value.
The phonon spectrum is calculated using phonopy software interfaced with VASP~\cite{phonopy}.
To study different magnetic structures, a 2$\times$2$\times$2 (2$\times$2) supercell for CrSBr bulk (monolayer) has been used in our calculations. The kinetic energy cutoff for plane wave expansion is set to 450 eV. The Monkhorst-Pack grid of 11$\times$9$\times$3 (11$\times$9$\times$1) k-mesh is used for CrSBr bulk (monolayer).
The setup of kinetic energy and k-point sampling is carefully tested to ensure the convergence of the MA results within about 2 $\mu$eV/Cr, see Supplemental Material~\cite{SM}.
To describe the correlation effect of Cr $3d$ electrons, the GGA plus Hubbard $U$ (GGA+$U$) is employed~\cite{U-method}, with the common values of $U$ = 4 eV and Hund
exchange $J_{\rm H}$ = 0.9 eV. This is very similar to the previous choices of $U$ = 4 eV and $J_{\rm H}$ = 1 eV (and the effective $U_{\rm eff}$ = $U-J_{\rm H}$ = 3 eV)~\cite{Jiang2018,Wang2020,Chen2020,Guo2018}.
We also test $U$ = 3 eV and 5 eV (see Supplemental Material~\cite{SM}), and find that with the increasing $U$, the semiconducting gap is slightly enlarged (due to its charge transfer type being less sensitive to the $U$), the spin moments and exchange parameters have only some quantitative changes, but the estimated $T_{\rm C}$ of CrSBr monolayer has an insignificant change from 180 K to 172 K. We also note that unlike the direct Coulomb repulsion $U$, $J_{\rm H}$ is actually the difference of the energies of electrons with different spins or orbitals in the same atomic shell, and therefore $J_{\rm H}$ is almost not screened and not modified when going from an atom to a solid. It is almost a constant for a given element and is typically 0.8–1.0 eV for a $3d$ transition metal~\cite{Khomskii}.
The spin-orbit coupling (SOC) is included in our calculations to evaluate the MA, together with the counting of the magnetic dipole-dipole interaction. $T_{\rm C}$ of CrSBr monolayer is estimated using
Monte Carlo simulations on 8$\times$8$\times$1 spin matrix with periodic boundary condition.
During the simulation steps, each spin is rotated randomly in the three-dimensional space.
The spin dynamical process is studied by the classical Metropolis simulations~\cite{Metropolis}.

\section*{III. Results and Discussion}

\subsection*{CrSBr Bulk}

 \begin{figure}[t]
\includegraphics[width=8cm]{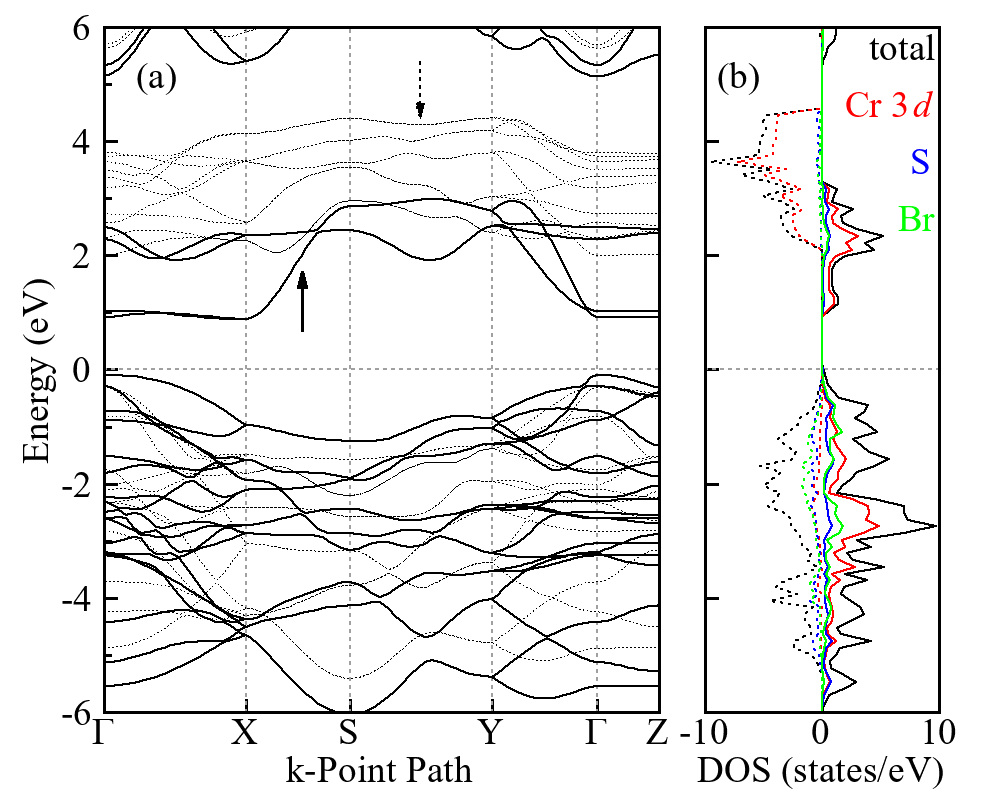}
 \caption{(a) The band structure of CrSBr bulk calculated by GGA+$U$. The solid (dashed) lines stand for the up (down) spin. The Fermi level is set at zero energy. (b) Density of states (DOS): the black curves stand for the total DOS, and the red, blue, and green curves refer to the Cr 3$d$, S, and Br contributions, respectively.
}
\label{Bulk_band}
\end{figure}

We first perform GGA+$U$ calculations for the CrSBr bulk in the FM state, and plot in FIG. \ref{Bulk_band} the band structure and the density of states (DOS). The interlayer AF coupling is calculated to be very weak (see below) and practically has no effect on the calculated electronic structures. The calculated band gap is about 1 eV, being well comparable to the experimental one of about 1.25 eV~\cite{Telford2020}. The main structural blocks, i.e., the distorted CrS$_4$Br$_2$ octahedra approximately yield the common crystal field splitting between the lower $t_{2g}$ triplet and higher $e_g$ doublet, both of which are however mixed due to the distortion.  In the energy range of 1-4 eV above the Fermi level, the two Cr ions in the FM $1\times$1$\times1$ cell contribute four unoccupied up-spin $e_g$ bands and ten empty down-spin $3d$ ($t_{2g}$+$e_g$) bands, indicating the formal Cr$^{3+}$ state with the $t_{2g}^3$ ($S$=3/2) configuration. From the DOS results, we see that the group of conduction bands lying in 1-4 eV is indeed composed mainly by the Cr $3d$ states. In strong contrast,  the topmost valence bands within 2 eV below the Fermi level have much less Cr $3d$ states. But they arise mainly from the S $3p$ and Br $4p$ states, both of which are fat orbitals and their DOS intensity is not well captured within their muffin-tin spheres but has a major distribution in the interstitial region. Therefore, the semiconducting band gap of bulk CrSBr can be identified to be the charge transfer type, which is typically the case for strongly covalent Mott insulators~\cite{ZSA_PRL}.
The formal Cr$^{3+}$ $S$=3/2 state is confirmed by the localized spin moment of 3.06 $\mu_{\rm B}$, see TABLE~\ref{tb1}. Moreover, the strong Cr-S and Cr-Br covalences induce a finite negative spin moment at the fat S and Br ions, being --0.12 $\mu_{\rm B}$ and --0.02 $\mu_{\rm B}$, respectively. Those individual spin moments plus the interstitial contribution sum up to the integer spin moment of 3 $\mu_{\rm B}$ per formula unit (fu).

 \begin{figure}[t]
\includegraphics[width=8cm]{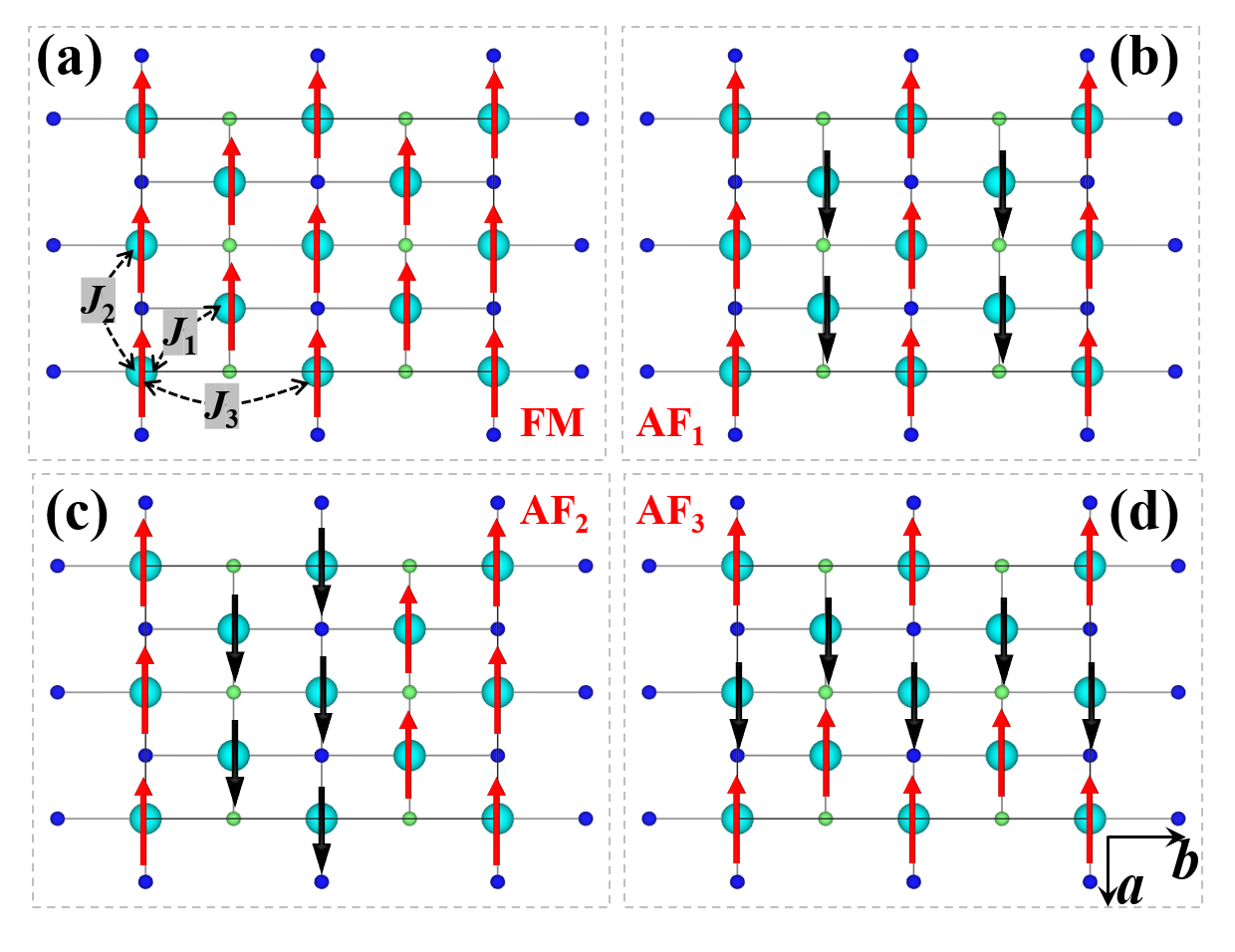}
 \caption{The four intralayer magnetic structures of CrSBr marked with three intralayer exchange parameters. Note that the interlayer magnetic coupling (not shown here) is three orders of magnitude weaker as seen in the main text.
}
\label{states}
\end{figure}

To study the magnetic properties of bulk CrSBr, we also carry out GGA+$U$ calculations for other three intralayer AF states besides the above FM state, see FIG.~\ref{states}. Our results show that the above intralayer FM solution is most stable and has a lower total energy than the three intralayer AF states by 27-35 meV/fu, see TABLE~\ref{tb1}. In addition, we calculate the interlayer AF state (with the intralayer FM) using the 2$\times$2$\times$2 supercell. We find, however, that the interlayer AF state differs from the FM state only by 0.04 meV/fu. These results suggest that the interlayer magnetic coupling is three orders of magnitude weaker than the intralayer FM exchanges, and thus it is marginal due to the vdW gap and of no more concern in this work. The major FM intralayer couplings, represented by the three FM exchange parameters $J_1$, $J_2$, and $J_3$ (see FIG.~\ref{states} and TABLE~\ref{tb1}), arise from the Cr-S-Cr and Cr-Br-S superexchanges, and they appear also in the CrSBr monolayer and will be discussed below.

\renewcommand\arraystretch{1.3}
\begin{table}[b]
\caption{Relative total energies $\Delta$\textit{E} (meV/fu), local spin moments ($\mu_{\rm B}$) and total spin moments ($\mu_{\rm B}$/fu) for the four intralayer magnetic structures of CrSBr bulk by GGA+$U$ (see also FIG.~\ref{states}). The derived three intralayer FM parameters (meV) are listed. Note that the interlayer magnetic coupling is three orders of magnitude weaker (interlayer AF and FM differing only by 0.04 meV/fu).
}

\begin{tabular}{c@{\hskip6mm}r@{\hskip6mm}r@{\hskip6mm}r@{\hskip6mm}r@{\hskip6mm}r@{\hskip6mm}}
\hline\hline
States  &$\Delta$\textit{E}&Cr$^{3+}$&S$^{2-}$&Br$^{-}$& \textit{M}$_{\rm tot}$ \\ \hline
FM      &  0            &  3.06  & --0.12 &      --0.02            &  3.00          \\
AF$_{1}$&  35          &  $\pm$2.97  &  $\mp$0.01  &       $\mp$0.01        &  0.00   \\
AF$_{2}$&  27         &   $\pm$3.02   &  $\mp$0.07    &        $\mp$0.01      &  0.00  \\
AF$_{3}$&  30         &  $\pm$2.96   &    $\mp$0.04     &   $\mp$0.00      &  0.00   \\
\hline
\multicolumn{2}{c}{\textit{J}$_{1}$=3.88}    & \multicolumn{2}{c}{\textit{J}$_{2}$=2.79} &    \multicolumn{2}{c}{\textit{J}$_{3}$=2.12}    \\
\hline\hline
 \end{tabular}
 \label{tb1}
\end{table}

Next, we study the experimental triaxial MA of bulk CrSBr. We address this issue first by including the SOC and carrying out GGA+$U$+SOC calculations. We assume the magnetization axis along the $a$, $b$, and $c$ axes, respectively, and then run their respective self-consistent calculations till the energy difference converges within 2 $\mu$eV/Cr. As seen in TABLE~\ref{tb2}, the SOC-MA has the easy $b$ axis, and the $a$ or $c$ axis magnetization has a higher energy by about 20 $\mu$eV/Cr. Then this weak SOC-MA seems uniaxial. The present $b$-axis easy magnetization is in line with the experiment, but the measured triaxial MA (easy $b$-axis, intermediate $a$-axis, and hard $c$-axis with the respective saturation magnetic fields of 0.58, 1.17, and 2.00 Tesla~\cite{Telford2020}) is not yet well reproduced.

\begin{table}[t]
\caption{The SOC-MA, shape-MA, and total-MA ($\mu$eV/Cr) for CrSBr bulk in the FM state.
}
\label{tb2}
\begin{tabular}{l@{\hskip9mm}c@{\hskip9mm}c@{\hskip9mm}c@{\hskip9mm}}
\hline\hline
Bulk            &010 (\textit{b})&100 (\textit{a}) &001 (\textit{c})  \\
\hline
SOC-MA         &0              &19              &21                \\
shape-MA       &0              &--3             &13                \\
total-MA             &0              &16              &34                \\
\hline\hline
 \end{tabular}
\end{table}

Considering this weak SOC-MA and the vdW layered structure, one may think of an MA contribution from the shape anisotropy. The shape MA originates from a magnetic dipole-dipole interaction, which is expressed as follows
\begin{equation}
E^{\rm dipole-dipole}=\frac{1}{2}\frac{\mu_0}{4\pi}\sum_{i \neq j}^{N} \frac{1}{r^{3}_{ij}}[\vec{M}_{i} \cdot \vec{M}_{j} -
\frac{3}{r^{2}_{ij}}(\vec{M}_{i} \cdot \vec{r}_{ij})(\vec{M}_{j} \cdot \vec{r}_{ij})],
\label{eq1}
\end{equation}
where $\vec{M}_{i}$ represents the Cr$^{3+}$ magnetic moments and $\vec{r}_{ij}$ is a vector connecting the Cr-sites \textit{i} and \textit{j}. In particular, when $\vec{M}_{i}$ and $\vec{M}_{j}$ are parallel ($i.e.,$ in a collinear FM structure) and have a same value, this equation is simplified as
\begin{equation}
E^{\rm dipole-dipole}=\frac{1}{2}\frac{\mu_0 M^{2}}{4\pi}  \sum_{i \neq j}^{N} \frac{1}{r^{3}_{ij}}[1 -
3\cos^{2}\theta_{ij}],
\label{eq2}
\end{equation}
where $\theta_{ij}$ is the angle between the $\vec{M}$ and $\vec{r}_{ij}$.

We list the shape-MA results in TABLE~\ref{tb2}, which show that the magnetization prefers to lie in the $ab$ plane as expected for the layered structure. The shape-MA is at the same energy scale of 10 $\mu$eV/Cr as the SOC-MA. Then we sum up the SOC-MA and shape-MA, and get the total MA: the $b$-axis is indeed the easy one, the $a$-axis is intermediate with the MA energy of 16 $\mu$eV/Cr relative to the easy $b$-axis, and the $c$-axis is hard with the MA energy of 34 $\mu$eV/Cr. Thus, our present results well reproduce the experimental triaxial MA~\cite{Telford2020}, and the MA energy values scale with the intervals among the experimental saturation magnetic fields of 0.58, 1.17, and 2.00 Tesla along the $b$, $a$, and $c$ axes, respectively~\cite{Telford2020}. To summarize, we find that bulk CrSBr is a magnetic semiconductor with the major FM intralayer couplings and the charge-transfer type band gap. Its triaxial MA is ascribed to the joint effect of SOC-MA and shape MA. All these results well account the experimental observations~\cite{Telford2020}.

\subsection*{CrSBr Monolayer}

 \begin{figure}[b]
\includegraphics[width=8cm]{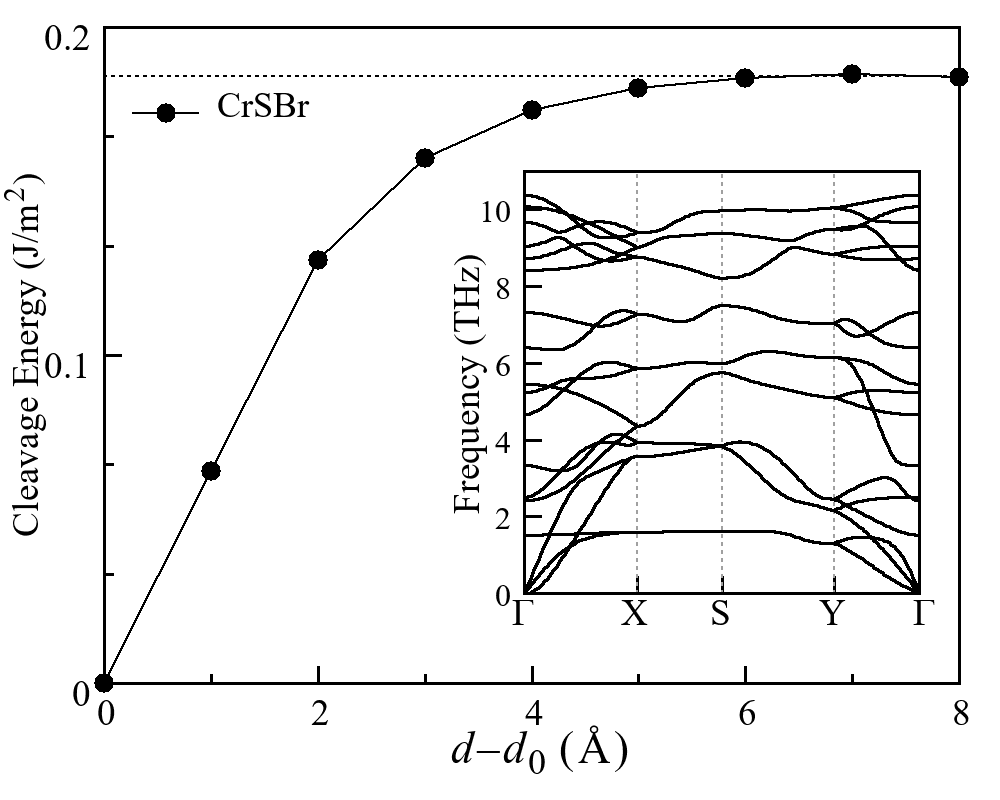}
 \caption{The relative energy as a function of the distance between two
CrSBr monolayers with respect to the experimental vdW distance \textit{d}$_{0}$, and
the phonon spectrum shown in the inset. There are no imaginary frequency over the Brillouin zone.
}
\label{exfoliation}
\end{figure}

CrSBr is a vdW layered material and has a major FM intralayer couplings, and therefore, its monolayer could be a 2D FM semiconductor. A structural optimization finds that CrSBr monolayer would have almost the same planar lattice constants (within 1\%) as the experimental bulk value. The calculated cleavage energy of 0.18 J/m$^{2}$ for the CrSBr monolayer (see FIG.~\ref{exfoliation}) is even smaller than that of 0.3 J/m$^{2}$ for the CrI$_{3}$ monolayer, which is already exfoliated from its bulk~\cite{Mcguire2015}. The dynamical stability of CrSBr monolayer is verified by the phonon calculations, which show no imaginary frequency phonons throughout the Brillouin zone, see the inset of FIG.~\ref{exfoliation}. Hence, we now turn to the CrSBr monolayer and study its electronic and magnetic properties using first-principles calculations and Monte Carlo simulations, aided by the analyses of the FM superexchange and triaxial MA.

 \begin{figure}[t]
\includegraphics[width=8cm]{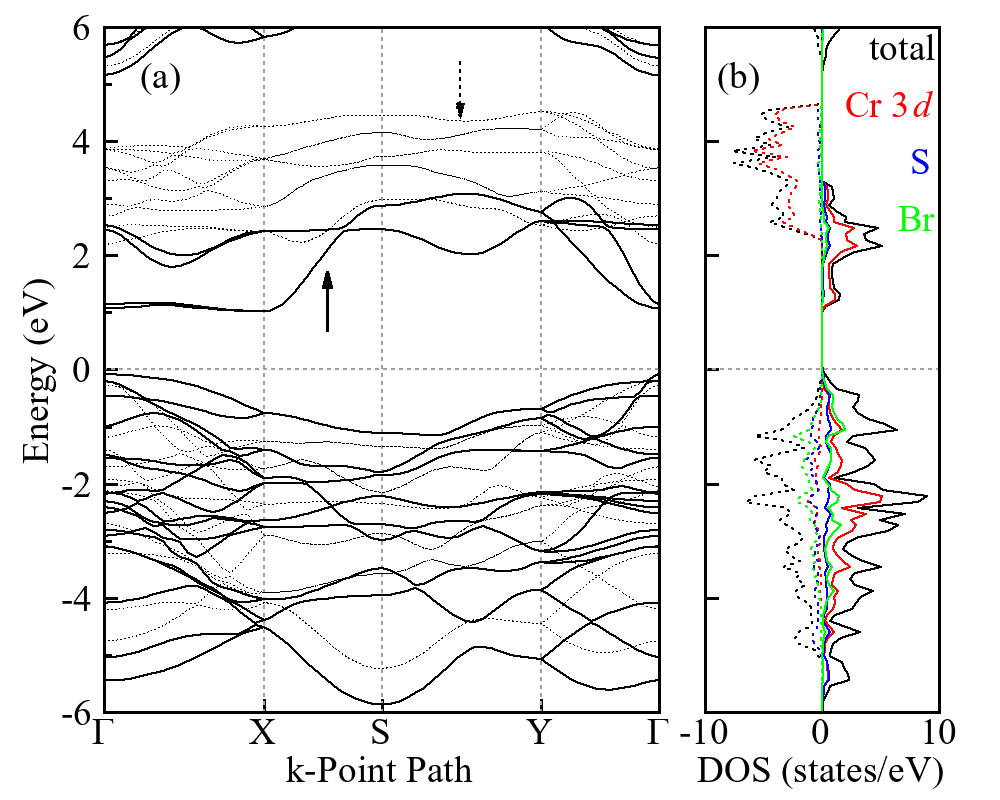}
 \caption{The band structure and DOS for CrSBr monolayer by GGA+$U$. See FIG.~\ref{Bulk_band} for a comparison.
}
\label{monolayer_band}
\end{figure}

The electronic band structure of CrSBr monolayer (see FIG.~\ref{monolayer_band}) is very similar to that of the bulk material (see FIG.~\ref{Bulk_band} for a comparison). The semiconducting band gap of the charge transfer type gets a little bigger for the monolayer, which is due to a little band-narrowing associated with the dimensionality reduction. We now check the intralayer magnetic couplings by studying the four magnetic states plotted in FIG.~\ref{states}. We define three exchange parameters, $J_1$ along the $ab$ diagonal direction for the SS-edge sharing CrS$_4$Br$_2$ octahedra, $J_2$ along the $a$ axis for the SBr-edge sharing CrS$_4$Br$_2$ octahedra, and $J_3$ along the $b$ axis for the Br-corner sharing CrS$_4$Br$_2$ octahedra, see Figs. 1 and 3. Counting $-JS^2$ for each pair of Cr$^{3+}$ $S$=3/2 ions (positive $J$ refers to FM exchange), the magnetic exchange energies of the FM state and the three AF states per formula unit are written as follows
\begin{equation*}
\begin{split}
E_{\rm FM} &=  (-2J_{1}-J_{2}-J_{3})S^{2},\\
E_{\rm AF_{1}} &= (2J_{1}-J_{2}-J_{3})S^{2},\\
E_{\rm AF_{2}} &= (-J_{2}+J_{3})S^{2},\\
E_{\rm AF_{3}} &= (J_{2}-J_{3}) S^{2}.
\end{split}
\end{equation*}
Again, we find that the FM solution is the ground state, as seen in TABLE~\ref{tb3}. It has the formal Cr$^{3+}$ $S$=3/2 state and a finite negative spin moment at the S and Br ions induced by the strong Cr-S and Cr-Br covalencies. Using the values of the relative total energies in TABLE~\ref{tb3} and the above equation, we can derive the three exchange parameters as listed in TABLE~\ref{tb3}, all of which are positive and thus of FM type.

According to the Goodenough-Kanamori-Anderson rules~\cite{Anderson1950,Goodenough1958,Kanamori1959}, the superexchange between the Cr$^{3+}$ ions with the $t_{2g}^3$ $S$=3/2 configuration is FM for a 90$^\circ$ bond angle, and it is AF for a 180$^\circ$ bond angle. Then it is natural that the near-90$^\circ$ Cr-S-Cr superexchange in the SS-edge sharing Cr$S_4$Br$_2$ octahedra gives the FM $J_1$, which is the case for the extensively studied CrI$_3$~\cite{Wang_EPL}. This also holds true for the near-90$^\circ$ Cr-S(Br)-Cr superexchange which yields the FM $J_2$. In contrast, $J_3$ arises from the Cr-S-Cr superexchange in the S-corner sharing Cr$S_4$Br$_2$ octahedra, and the Cr-S-Cr bong angle of 160$^\circ$ would give competing AF and FM interactions.
In addition, the local distortion of the Cr$S_4$Br$_2$ octahedra mixes up the otherwise orthogonal $t_{2g}$ and $e_g$ orbitals, and such a mixing would also introduce a FM contribution even in the 180$^\circ$ superexchange.
Eventually, $J_3$ = 2.78 meV turns out to be FM in the present case, but it is smaller than the FM $J_1$ = 4.11 meV and $J_2$ = 3.22 meV due to its reduction by the competing AF interaction. Moreover, the increase (decrease) of $J_3$ upon the compressive (tensile) strain along the $b$ axis signals more (less) effective near-90$^\circ$ FM superexchange (see below), which is also in line with the present discussion. So far, we have seen that CrSBr monolayer is a charge-transfer type semiconductor and has the major FM intralayer couplings.

\renewcommand\arraystretch{1.3}
\begin{table}[t]
\caption{Relative total energies $\Delta$\textit{E} (meV/fu), local spin moments ($\mu_{\rm B}$) and total spin moments ($\mu_{\rm B}$/fu) for the CrSBr monolayer by GGA+$U$. The derived three exchange parameters (meV) are listed.}
\label{tb3}
\begin{tabular}{c@{\hskip6mm}r@{\hskip6mm}r@{\hskip6mm}r@{\hskip6mm}r@{\hskip6mm}r@{\hskip6mm}}
\hline\hline
States  &$\Delta$\textit{E}&Cr$^{3+}$&S$^{2-}$&Br$^{-}$& \textit{M}$_{\rm tot}$  \\ \hline
FM      &  0            &  3.07     &   --0.13  &   --0.02       &  3.00            \\
AF$_{1}$&  37         &  $\pm$2.98     &    $\mp$0.01  &    $\mp$0.02     &  0.00     \\
AF$_{2}$&  31         &  $\pm$3.02     &    $\mp$0.07  &     $\mp$0.02   &  0.00       \\
AF$_{3}$&  33          &  $\pm$2.98      &   $\mp$0.05   &     $\mp$0.00   &  0.00  \\
\hline
\multicolumn{2}{c}{\textit{J}$_{1}$=4.11}    & \multicolumn{2}{c}{\textit{J}$_{2}$=3.22} &    \multicolumn{2}{c}{\textit{J}$_{3}$=2.78}    \\
\hline\hline
 \end{tabular}
\end{table}

We now address the MA of CrSBr monolayer, which is of concern for a potential 2D FM semiconductor. Here we consider again the SOC-MA and shape-MA. For the monolayer with a collinear FM, Eq.~\ref{eq2} holds when the spins lie in the plane
\begin{equation}
E^{\parallel}=\frac{1}{2}\frac{\mu_0 M^{2}}{4\pi}  \sum_{i \neq j}^{N} \frac{1}{r^{3}_{ij}}[1 -
3\cos^{2}\theta_{ij}].
\label{eq3}
\end{equation}
However, when the spins are out-of-plane, $\theta_{ij}$=90$^\circ$ applies for all sites \textit{i} and \textit{j}, and Eq.~\ref{eq2} can further be simplified as
\begin{equation}
E^{\perp}=\frac{1}{2}\frac{\mu_0 M^{2}}{4\pi}  \sum_{i \neq j}^{N} \frac{1}{r^{3}_{ij}}.
\label{eq4}
\end{equation}
Then, the shape-MA is now expressed as
\begin{equation}
E^{\rm \perp} - E^{\rm \parallel} =\frac{3}{2}\frac{\mu_0 M^{2}}{4\pi}  \sum_{i \neq j}^{N} \frac{1}{r^{3}_{ij}}\cos^{2}\theta_{ij}.
\label{eq5}
\end{equation}
From Eqs.~1-5, one can expect that with the decreasing dimensionality of the materials, the shape-MA (out-of-plane against in-plane) gradually increases till it reaches the maximum at the monolayer. The shape-MA favors an in-plane magnetization and it plays an important role for low-dimensional magnetic materials.

We summarize in TABLE~\ref{tb4} the calculated SOC-MA and shape-MA for CrSBr monolayer.
The SOC-MA results show that the $b$-axis is easy one but the $a$-axis would be hard, instead of the hard $c$-axis for the bulk material. As for the shape-MA, the $c$-axis magnetization energy is much higher than those along the $a$ and $b$ axes as expected, showing a pronounced contribution of shape-MA for this 2D FM. Summing up the SOC-MA and shape MA, the total MA gives the easy magnetization $b$-axis, intermediate $a$-axis, and hard $c$-axis for CrSBr monolayer. Such a magnetic behavior is the same as the experimental triaxial MA observed for the bulk.

\begin{table}[t]
\caption{The SOC-MA, shape-MA, and total-MA ($\mu$eV/Cr) for CrSBr monolayer in the FM ground state.
}
\label{tb4}
\begin{tabular}{l@{\hskip9mm}c@{\hskip9mm}c@{\hskip9mm}c@{\hskip9mm}}
\hline\hline
Monolayer           &010 (\textit{b})&100 (\textit{a}) &001 (\textit{c})  \\
\hline
SOC-MA         &0              &22              &11                \\
shape-MA       &0              &--10            &67                \\
total-MA             &0              &12              &78                \\
\hline\hline
 \end{tabular}
\end{table}
 \begin{figure}[b]
\includegraphics[width=7cm]{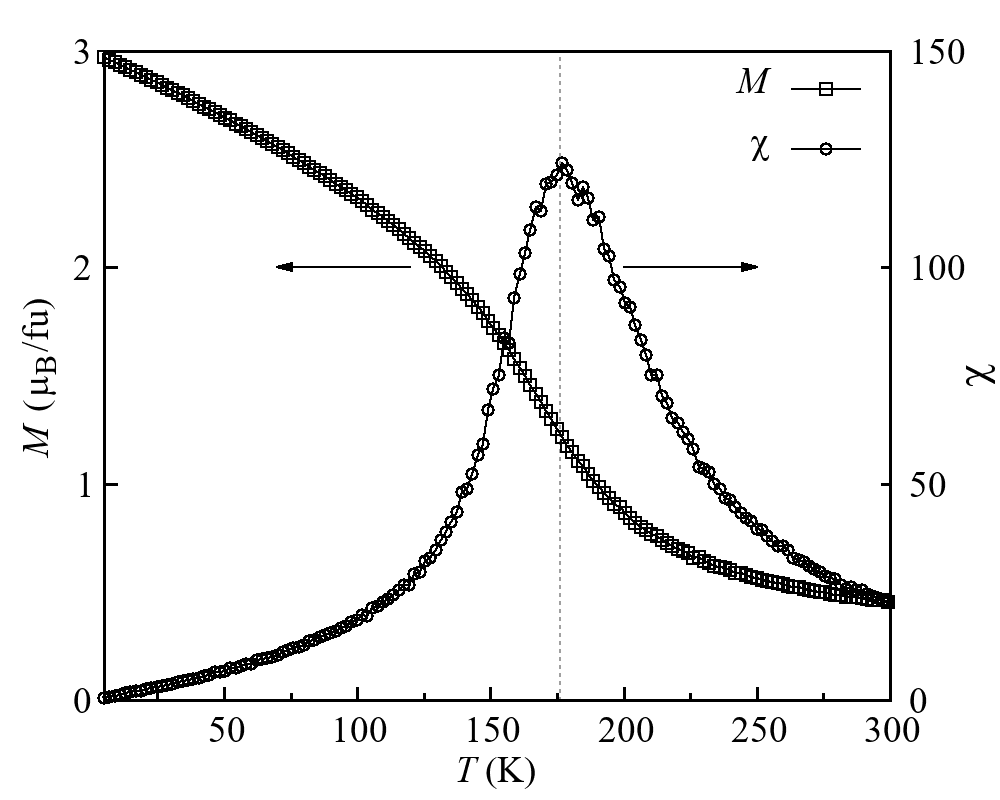}
 \caption{Monte Carlo simulations of the magnetization and magnetic susceptibility for CrSBr monolayer.
}
\label{MC}
\end{figure}

With the above study of the FM intralayer couplings and the triaxial MA, we now assume the following spin Hamiltonain and carry out Monte Carlo simulations to estimate $T_{\rm C}$ for CrSBr monolayer
\begin{align}
\begin{split}
H = & \sum_{k=1,2,3}-\frac{J_{k}}{2} \sum_{i,j}{\overrightarrow{S_{i}} \cdot \overrightarrow{S_{j}}} \\
    &    + D\sum_{i}{(S_{i}^{c})^{2}} + E_n\sum_{i}((S_{i}^{a})^{2}-(S_{i}^{b})^{2}),
\label{eq6}
\end{split}
\end{align}
where the first term describes the isotropic Heisenberg exchange, and the single ion anisotropy parameters $D$ and $E_n$ are introduced to simulate the triaxial MA. Then using the results listed in Tables~\ref{tb3} and \ref{tb4}, our Monte Carlo simulations find that $T_{\rm C}$ is 175 K for CrSBr monolayer (FIG.~\ref{MC}), which agrees with the previous predictions~\cite{Jiang2018,Wang2020,Chen2020,Guo2018}.

 \begin{figure}[t]
\includegraphics[width=8cm]{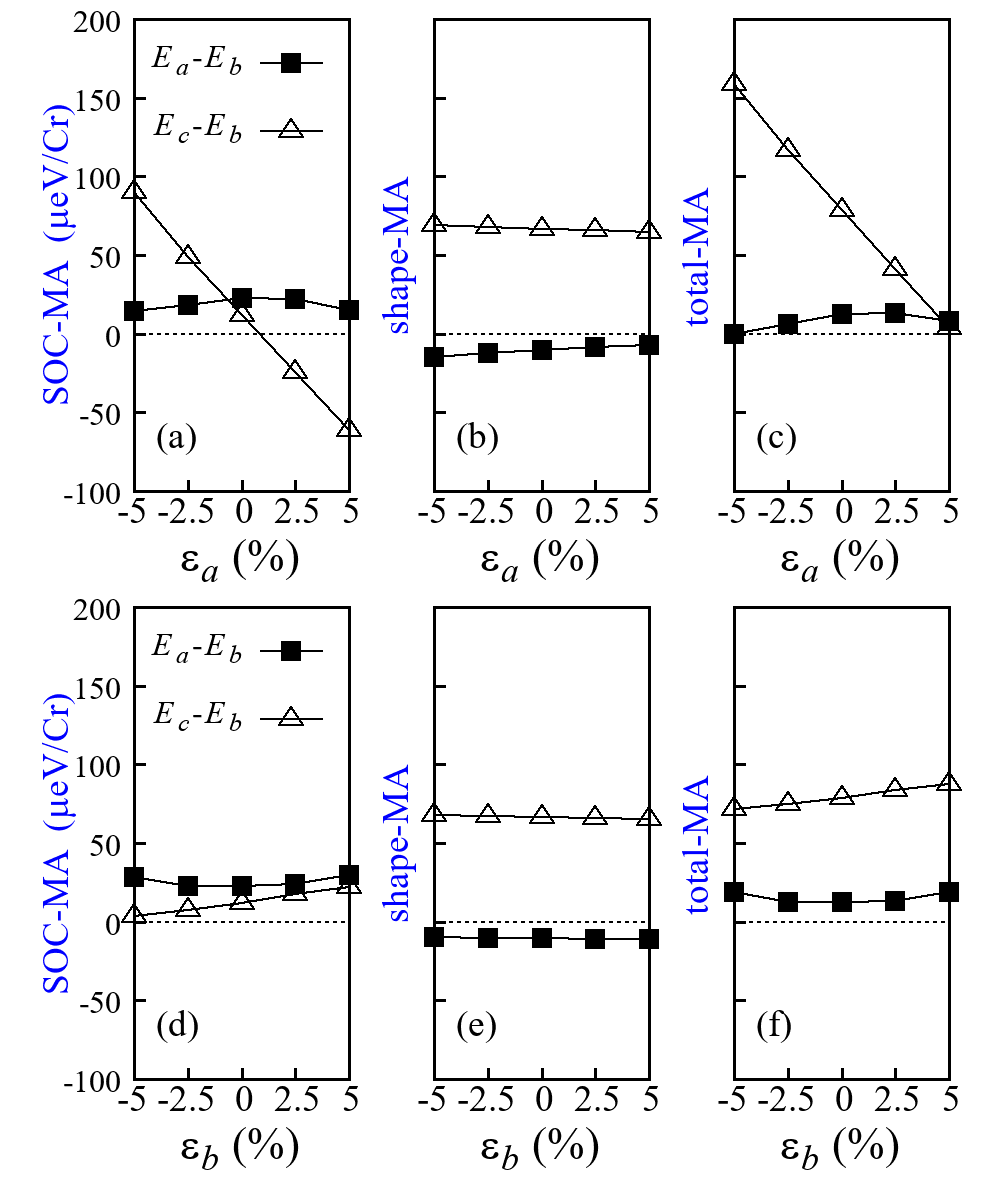}
 \caption{The SOC-MA, shape-MA, and total-MA ($\mu$eV/Cr) for CrSBr monolayer under a uniaxial strain along the $a$ or $b$ axis.
}
\label{strain-MA}
\end{figure}

We also check the strain effects of CrSBr monolayer and find that for the uniaxial strain along the $a$-axis, the $c$-axis MA energy induced by the SOC changes rapidly and even becomes negative upon a tensile strain, see FIG.~\ref{strain-MA}(a). In this case, the shape MA changes little, see FIG. \ref{strain-MA}(b). For the uniaxial strain along the $b$-axis, however, both the SOC-MA and shape MA undergo insignificant changes, see Figs.~\ref{strain-MA}(d)-(f). The total MA results show that the easy $b$-axis persists upon the strain of $\pm$5\%.

\begin{figure}[t]
 \centering
\includegraphics[width=7cm]{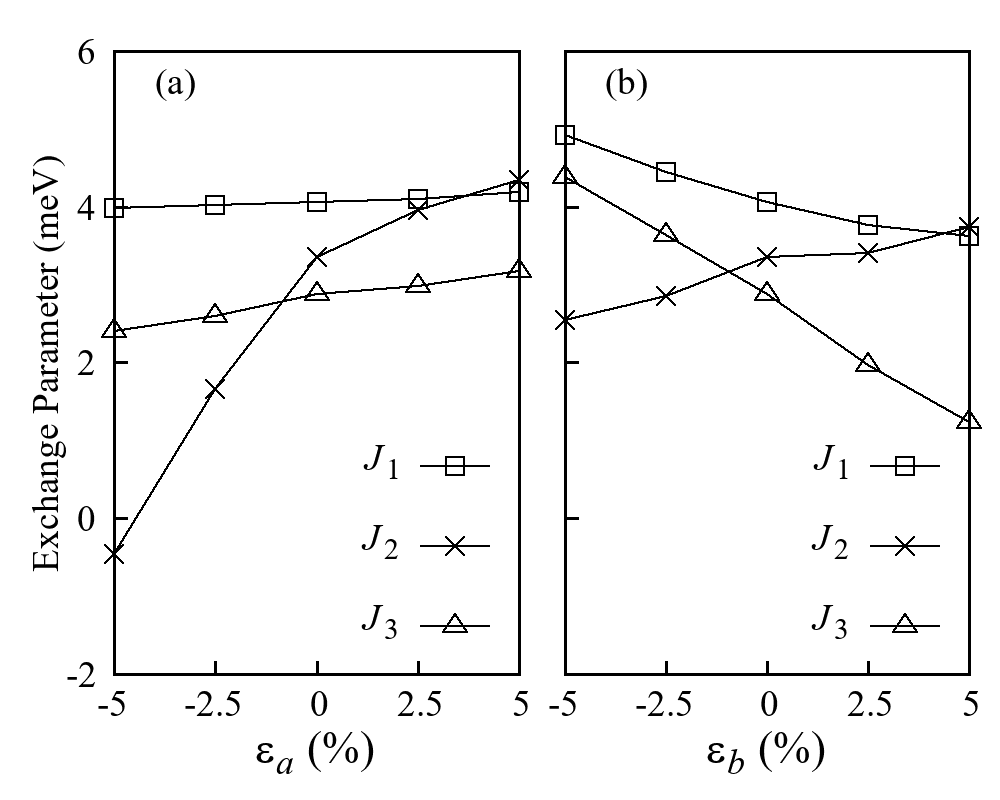}
 \caption{The three FM exchange parameters for CrSBr monolayer under a uniaxial strain along the $a$ or $b$ axis.
}
\label{strain-J}
\end{figure}

\begin{figure}[H]
\includegraphics[width=7cm]{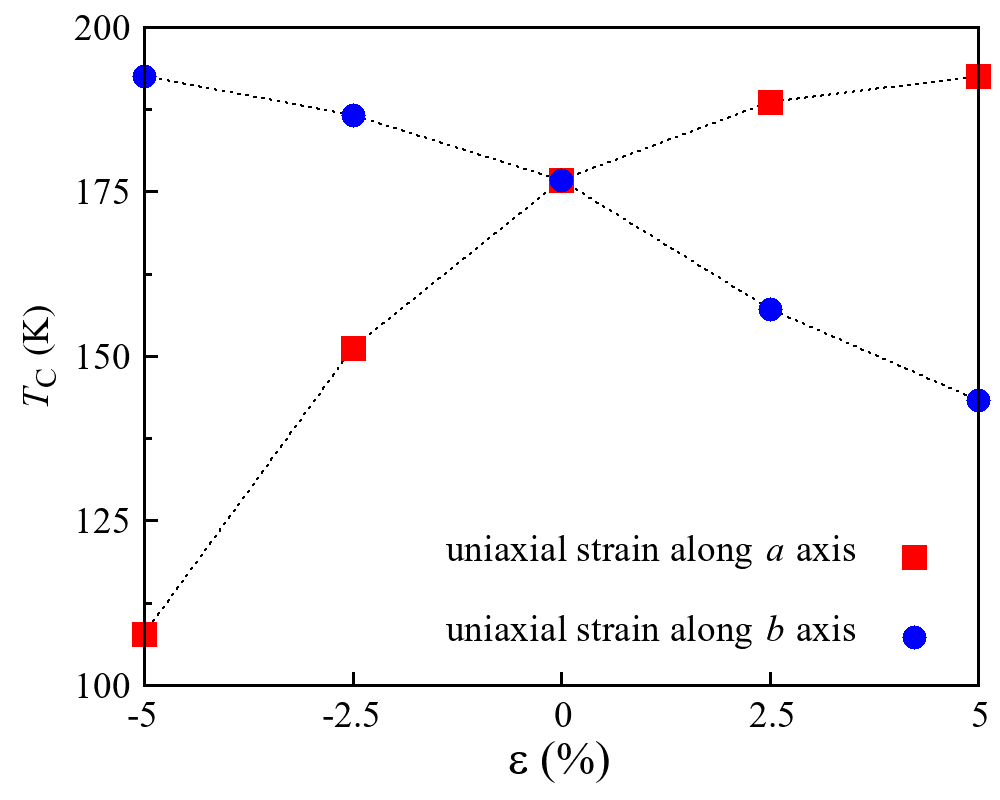}
 \centering
 \caption{$T_{\rm C}$ of CrSBr monolayer under a uniaxial strain along the $a$ or $b$ axis.
}
\label{TC}
\end{figure}

As for the three FM exchange parameters, we find that $J_2$ is moderately enhanced by a tensile strain along the $a$-axis but strongly reduced by a compressive strain, and that $J_3$ is enhanced (reduced) by a compressive (tensile) strain along the $b$-axis, see FIG.~\ref{strain-J}. Then it is expected that $T_{\rm C}$ of CrSBr monolayer would be increased by  a tensile strain along the $a$-axis or by a compressive strain along the $b$-axis, which is indeed confirmed by our Monte Carlo simulations, see FIG.~\ref{TC}.


\section*{IV. Summary}

To conclude, we find using density functional calculations that the vdW layered material CrSBr is a charge-transfer type semiconductor and has the major FM intralayer superexchange via the strongly covalent Cr-S(Br)-Cr bonds. Our calculations including the SOC-MA and shape MA well reproduce the experimental easy $b$-axis, intermediate $a$-axis, and hard $c$-axis. Moreover, we find that the FM superexchange and the triaxial MA persist in CrSBr monolayer, and that the shape MA becomes more pronounced in this 2D FM semiconductor. Using those FM exchange parameters and the triaxial MA, our Monte Carlo simulations yield $T_{\rm C}$ = 175 K, which can further be enhanced by a tensile strain along the $a$-axis or by a compressive strain along the $b$-axis.

\section*{Acknowledgements}
This work was supported by National Natural Science
Foundation of China (Grants No. 12104307 and No.
12174062) and by the National Key Research and Development
Program of China (Grant No. 2016YFA0300700).


\bibliographystyle{apsrev4-1}
\bibliography{CrSBr}

\end{document}